\documentclass{elsart}
\usepackage{amssymb}
\usepackage{epsfig,epsf,graphicx,amsmath}
\usepackage{color}
\begin{document}
\begin{frontmatter}
\title{ Core momentum distribution in two-neutron halo nuclei }
\author[ITA]{L. A. Souza},
\author[ITA,AARHUS]{F. F. Bellotti},
\author[IFT]{M. T. Yamashita},
\author[ITA]{T. Frederico} 
\footnote{Corresponding author: tobias@ita.br}
and
\author[IFT,UFABC]{Lauro Tomio}
\address[ITA]{Instituto Tecnol\'ogico de Aeron\'autica, DCTA,
12228-900, S. Jos\'e dos Campos,~Brazil.}
\address[AARHUS]{Department of Physics and Astronomy, Aarhus University, DK-8000 Aarhus C, Denmark.}
\address[IFT]{Instituto de F\'\i sica Te\'orica, UNESP - Universidade Estadual Paulista,  
01156-970, S\~ao Paulo, Brazil}
\address[UFABC]{Centro de Ci\^encias Naturais e Humanas, 
Universidade Federal do ABC, 09210-580, 
Santo Andr\'e, Brazil.}
\date{\today}
\maketitle
\begin{abstract}
The core momentum distribution of a weakly-bound 
neutron-neutron-core exotic nucleus is computed 
within a renormalized zero-range three-body model, 
with interactions in the s-wave channel. The
halo  wave-function in momentum space is obtained by using as inputs the 
two-body scattering lengths and the two-neutron 
separation energy. The core momentum densities are computed for $^{11}$Li, $^{14}$Be
$^{20}$C and $^{22}$C. The model describes the experimental data for $^{11}$Li, $^{14}$Be 
and to some extend $^{20}$C. The recoil momentum  distribution  of  the $^{20}$C 
 from the breakup of $^{22}$C  nucleus is computed for 
 different two-neutron separation energies, and
 from the comparison with recent experimental data the  two-neutron separation 
 energy  is estimated in the range
$100\lesssim S_{2n}\lesssim 400$ KeV. The recoil momentum distribution depends 
weakly on the neutron-$^{20}$C 
scattering length, while the  matter radius is strongly sensitive to it.
The expected universality of the momentum distribution width is verified by 
also considering  excited states for the system. 
\end{abstract}
\vspace{-0.5cm}
\begin{keyword}
Binding energies, Faddeev equation, halo-nuclei, three-body system
\end{keyword}
\end{frontmatter}

\section{Introduction}

The core recoil momentum distribution of radioactive two-neutron
halo nuclei close to the drip line, extracted from breakup reactions 
at few hundreds MeV/A, are expected to be quite useful in order to 
get insights on the underlying neutron-neutron-core structure of these
exotic nuclei~\cite{TanJPG96,RiiPST13}. This is particularly clear in the example of 
$^{11}$Li breakup in a carbon target at 800 MeV/A~\cite{TanJPG96}, where 
the momentum distribution is characterized by tahe sum of two distributions, 
a narrow one with   $\sigma=21(3)$MeV/c and a wide one with $\sigma= 80$ MeV/c, 
given that $\sigma^2$ is the variance associated with a normal distribution.
The narrow momentum distribution should be associated with a large configuration of 
the two neutrons forming a halo structure. In this case the breakup occurs 
when the two neutrons 
are found quite far from the core, corresponding to a  weakly-bound three-body
system in the nuclear scale.
On the other hand, the wide momentum distribution is related to the 
inner part of the halo neutron orbits, close to the core region. 

An interesting aspect of two-neutron halo states, associated with the narrow core 
momentum distribution, is that the halo constituents should have a high probability to be 
found in the classically forbidden region, outside the potential range. Therefore, the halo wave function 
should be quite insensitive to details of the interactions, once the model is adjusted by the best known
two- and three-body low-energy observables. Therefore, one natural observable is the 
the two-neutron separation energy, $S_{2n}$, which represent the three-body binding. For the 
two-body subsystems neutron-neutron $(n-n)$ and neutron-core $(n-c)$, the appropriate 
observables are the corresponding scattering lengths (or, respective, two-body energies). 
With these arguments, studies with schematic potentials, such as contact interactions, have been 
quite successful in describing low-energy three-body structures for large two-body scattering lengths 
(when the corresponding energies are close to zero). Actually, investigations on quantum three-body 
systems within this regime, in nuclear and atomic physics, became quite well known in view of recent 
experimental realizations in atomic laboratories of the long-time predicted Efimov 
effect~\cite{Efimov70}, which corresponds to the increasing number of excited three-body states as
one goes to the unitary limit (when one or both two-body scattering lengths are  close to infinity).
For recent reports, quoting the main experimental realizations of this effect, see 
Refs.~\cite{Efimov11,Ferlaino11}.

By considering a contact interaction, the corresponding wave-function is an eigenstate of the free 
Hamiltonian, except in the positions where the particles are right on the top of each other;
and, therefore, the particles are in the classically forbidden region (see, e.g., \cite{FreFBS14}). 
Such theoretical approach applied to light-exotic nuclei close to the neutron drip-line,
within a neutron-neutron-core ($n-n-c$) configuration, is described in detail in a recent review,  
in Ref.~\cite{FrePPNP12},  where universal aspects of the properties of the weakly-bound 
$n-n-c$ systems are  emphasized. 

Our focus here is to present a theoretical investigation concerned to the  
experimental core recoil momentum distributions of the halo-nuclei 
$^{11}$Li~\cite{TanJPG96}, $^{14}$Be~\cite{ZahPRC93}, $^{20}$C and $^{22}$C~\cite{KobPRC12}, 
as obtained by the halo breakup on nuclear targets (see also Ref.~\cite{TanPPNP12}). The approach
is the above described three-body model, which we found appropriate for the analysis of low-binding
energy systems as these ones. 
In the particular cases of $^{11}$Li, $^{14}$Be and the carbon systems $^{20}$C and $^{22}$C, 
we consider that the neutron-neutron and the neutron-core interactions are dominated by $s-
$wave states. 
The calculations of core momentum distributions are performed within a renormalized zero-
range three-body model, with the halo nucleus described as two neutrons 
with an inert core ($n-n-c$)~\cite{FrePPNP12,FreFBS11}. 
The detailed expressions for the momentum distribution are given in~\cite{YamPRA13}, within 
an approach that
requires as inputs one two-body ($n-c$) and one three-body ($n-n-c$ ) observable, 
given that the other two-body observable is fixed to the well-known virtual-state 
energy of the $n-n$ system. Usually, within such 
approach it is appropriate to consider the corresponding two-body scattering lengths 
(positive, for bound, and 
negative for virtual state systems); with a three-body scale given by the
 two-neutron separation energy,  $S_{2n}$.
Therefore, in a more general description of low-energy three-body physics 
with two distinguished particles
($\alpha-\alpha-\beta$), an appropriate universal scaling function is
 given (see e.g. \cite{FrePPNP12}), where only three 
low-energy inpus are enough to determine any other relevant low-energy
 observable of the system. 

Within our study on the momentum distributions of the core in halo nuclei the observable
that we are concerned is the variance of the momentum distribution, given by $\sigma^2$
(associated with the normal one), 
which is universally correlated to the two possible scattering lengths and $S_{2n}$. 
One obtains $\sigma$ from the Full Width at Half Maximum (FWHM) of the momentum 
distribution, such that on can find that FWHM $=2\sqrt{2\ln 2}\,\sigma$.
Once this quantity is known experimentally, one can use the scaling function to estimate the 
value of $S_{2n}$ or, eventually, to constraint some other poorly known low-energy observable, 
such as a subsystem energy, or scattering length.
The natural units for $\sigma$ in halo physics is MeV/c.
As we are interested in scaling properties of 
observables, it is convenient to introduce the dimensionless 
ratio $\sigma/\sqrt{S_{2n}m_n}$, where 
$m_n$ is the neutron mass. By taking $m_n$ as the mass unit, a scaling function can be 
defined, with a general form given by 
\begin{equation}
\frac{\sigma}{\sqrt{S_{2n}}}\,=\, {\cal S}_c\left(\pm\sqrt{\frac{E_{nn}}{S_{2n}}},\,
\pm\sqrt{\frac{E_{nc}}{S_{2n}}}; {A}\right) \, 
,\label{sigmac}
\end{equation}
where the $+$ and $-$ signs refer to the bound and virtual subsystem energies, respectively. 
The core mass number is $A\equiv m_c/m_n$. The corresponding 
energies, $E_{nn}$ and $E_{nc}$, are positive defined quantities, 
with $a_{nn}$ and $a_{nc}$ being the 
respective two-body scattering lengths.
In our specific case of the two-neutron halo nuclei the above scaling 
function (\ref{sigmac}) has  $E_{nn}$ fixed to the $n-n$ virtual state.
In the next, our units are such that the Planck constant $\hbar$ and the velocity 
of light $c$ are set to one. All masses are taken in units of $m_n$.

For the momentum distribution width, the scaling function (\ref{sigmac}) is the limit cycle 
of the correlation 
function associated with $\sigma$ as a function of $E_{nn}$, $E_{nc}$ and $S_{2n}$, when the 
three-body 
ultraviolet (UV) cut-off is driven to infinite in the three-body integral 
equations, or equally the 
scattering lengths driven to zero with a fixed UV cut-off. Similar procedure is 
performed within a renormalized zero-range three-body model, 
in the subtracted integral equations, where the subtraction energy is fixed and the
two-body scattering lengths are driven towards  infinite. 
In practice, both procedures provides very close results, as shown in Ref.~\cite{YamPRA02}.
In the exact Efimov limit ($E_{nn}=E_{nc}=0$), the width is a universal function of the 
mass number $A$, 
$\sigma/\sqrt{S_{2n}}\,=\, {\cal S}_c\left(0,\,0, { A}\right),$
which is associated to a limit cycle. Already in the first cycle it approaches 
the results of the
renormalized zero-range three-body model~(see e.g. \cite{FrePPNP12}), namely given by the 
subtracted 
Skorniakov and Ter-Martirosian equations for mass imbalanced systems~\cite{SKT}.

For the analysis of the core momentum distribution, we consider data 
for $^{11}$Li~\cite{TanJPG96}, 
$^{14}$Be~\cite{ZahPRC93} and  $^{20}$C~\cite{KobPRC12} as the low-energy parameters, 
which are 
the inputs of our renormalized zero-range model. This procedure allows us to 
verify the utility of such ``bare" formula (\ref{sigmac}), 
which does not include distortion effects from the scattering, to analyse the actual breakup 
data for those systems, taken at few-hundred MeV/A.  

As an application of our model, we study in more detail the two neutron halo of 
the Borromean nuclei $^{22}$C, in an attempt to extract information of the 
halo properties, by using the correlation between observables expressed in Eq.(\ref{sigmac}),
namely the width of the core recoil distribution as a function of 
$S_{2n}$ and the energy of the $s-$wave virtual state of $^{21}$C. 
From the experimental point of view
the two-neutron separation energy of $^{22}$C
is not well constrained, with a value of 0.42 $\pm$ 0.94 MeV given by systematics~
\cite{AudNPA03} 
and from a mass measurement, it was found $S_{2n}=$ -0.14(46) MeV\cite{GauPRL12}. 
There is an indirect evidence that 
$^{22}$C could be bound by less than 70 keV \cite{MosNPA13}. 
Other independent information on the
binding energy of this nucleus can be obtained from the matter radius. Tanaka and 
collaborators~\cite{TanPRL10} extracted a root-mean-square (rms) matter radius of 
$5.4\pm0.9$~fm from the analysis of  the large reaction cross sections of $^{22}$C on liquid 
hydrogen target at 40A MeV, using a finite-range Glauber calculation under an optical-limit 
approximation. 
Furthermore, the two-valence neutrons occupy preferentially one $s_{1/2}$ orbital 
in their analysis. Such rms matter radius, taken together with the corresponding 
one of $^{20}$C 
(2.98(5) fm\cite{OzaNPA01}), suggest a halo neutron orbit with rms radius of $15\pm4$~fm in 
$^{22}$C, which is 
constraining the $S_{2n}$ to be below 100 keV~\cite{YamPLB11}. 
This value is consistent with results obtained from a shell-model 
approach~\cite{ForPRC12} and 
results from effective field theory  with contact interaction~\cite{AchPLB13,AchFB15}. The 
estimated $^{22}$C quantities should be compared with the fairly small value of 
$S_{2n}=369.15(65)$ keV for $^{11}$Li  in the nuclear scale~\cite{SmiPRL08}, 
and with the neutron-neutron ($n-n$) average separation distances $R_{nn}$ 
in $^{11}$Li around 6-8 fm, which is obtained from the $n-n$ correlation 
function measured by the breakup cross-section 
on heavy nuclei \cite{MarPRC01,PetNPA04}. 
However, Riisager~\cite{RiiPST13} pointed out that a comparison of experimental
 data obtained for 
 the core recoil  momentum distributions of $^{11}$Li~\cite{TanJPG96} 
 and $^{22}$C~\cite{KobPRC12}  
 suggests similar neutron halo sizes for these nuclei, which could indicate an 
 overestimation of the 
 matter radius of this carbon isotope.

Our present work can give more insights in resolving the issue of the size of the
two neutron halo in  $^{22}$C. The constraints in the parameters associated with the $^{22}$C 
halo structure and 
two-neutron separation energy provided by the scaling formula for the width of the core 
recoil momentum distribution are discussed on the basis that corresponding data, 
fitted to three-body model calculations. The particular case of $^{22}$C is 
interesting 
considering that the corresponding observables are  probably dominated by the tail 
of the three-body wave function in an ideal $s-$wave three-body model. That ideal
 structure was already considered in Ref.~\cite{suzuki-2006}, within 
 a Borromean $n-n-^{20}$C configuration for  $^{22}$C, where all 
 two-body subsystems, $n-^{20}$C and $n-n$ are not bound.

As it will be shown in the following, the recent experimental results for $^{20}$C and $^{22}
$C \cite{KobPRC12} 
allow us, in principle, 
to constraint $S_{2n}$ and the matter radius  of $^{22}$C, even considering that the 
scattering length of the
subsystem neutron-$^{20}$C is not well known. From the experimental analysis
 performed in Ref.~\cite{MosNPA13},
the associated $s-$wave virtual-state energy of $^{21}$C is found to be about $1$ MeV. 
  
The present study on the constraint for $S_{2n}$ are relying on the applicability of the 
renormalized 
three-body zero-range  model and scaling function (\ref{sigmac}) derived for the width of the 
core recoil 
momentum distribution. In the case of  $^{22}$C, this is obtained by fitting this distribution to 
the experimental 
breakup cross-section data given in Ref.~\cite{KobPRC12}. For our estimative of $S_{2n}$ is 
also essential 
that the scaling function given in (\ref{sigmac}) has a weak dependence of 
the $E_{nc}/S_{2n}$ ratio.

One of the sources of information on the sizes of unstable neutron-rich nuclei, 
is the $n-n$ correlation function obtained from Coulomb breakup experiments with 
neutron rich projectile on heavy nuclei~\cite{MarPLB00,MarPRC01,PetNPA04}. 
The experimental results for the $n-n$ correlation function for Borromean 
nuclei $^{11}$Li and $^{14}$Be 
are found quite consistent with the corresponding computed quantities obtained within a 
subtracted renormalized zero-range 
model~\cite{YamPRC05}, unless an unexpected theoretical minimum before the correlation 
function approach 
unity for large relative momentum. Data from the experiments are showing a monodic decrease 
of the
correlation function with momentum; however, the accessible data goes only up to the 
predicted minima region.

Next, we present the basic formalism. In section 3 we have the main results, followed by the 
section 4 where we summarize our conclusions.

\section{Model formalism}
In the following, we briefly sketch the formalism, based on the renormalized zero-range three-body model,
leading to the core recoil momentum distribution formula, which is used in our data analysis of the halo
nuclei systems $^{11}$Li, $^{14}$Be,  $^{20}$C  and  $^{22}$C.
 
The renormalized zero-range model which we are considering to describe the halo wave-function has 
been explained in detail in the review \cite{FrePPNP12}. In order to built the $s-$wave three-body wave 
function for the $n-n-c$ system, one needs to solve a coupled integral equation for the independent 
spectator functions  $\chi_{nn}(q)$ and $\chi_{nc}(q)$. Within the zero range model, a regularization
is needed, which can be implemented with a cutoff momentum parameter, such as in Ref.~\cite{AmoPRC97},
or by considering the subtraction procedure used in ~\cite{YamNPA04}, which we follow in the 
present approach. Therefore, the present subtractive regularization approach for the spectator functions
is performed at a given energy scale $\mu^2$, by the following coupled equations:
{\small 
\begin{eqnarray}
\label{spec}
 \chi_{nc}({q})&=&\tau_{nc}(q;S_{2n})
\int_0^\infty k^2dk\left\{\left[ {\cal G}_1(q,k;S_{2n}) - {\cal G}_1(q,k;\mu^2) \right]\chi_{nc}(k) +
\right.\nonumber \\ &&\hspace{3.2cm}\left. 
+\left[ {\cal G}_2(q,k;S_{2n}) - {\cal G}_2(q,k;\mu^2) \right]\chi_{nn}(k)\right\},\\
\chi_{nn}({q})&=&2\tau_{nn}(q;S_{2n})
\int_0^\infty k^2dk \left[ {\cal G}_2(k,q;S_{2n}) - {\cal G}_2(k,q;\mu^2) \right]\chi_{nc}(k)
,\nonumber \end{eqnarray} 
}  where 
\begin{eqnarray}
\tau_{nc}(q;S_{2n})&\equiv &
 \sqrt{\left(\frac{A+1}{2A}\right)^3}\frac{1}{\pi} \left(\sqrt{S_{2n}+\frac{ 
A+2}{2A+2}\,q^2}\mp\sqrt{E_{nc}}\right)^{-1},\nonumber\\
\tau_{nn}(q;S_{2n})&\equiv &
 \frac{1}{\pi}\left(\sqrt{S_{2n}+\frac{A+2}
{4{ A}}q^2}\mp\sqrt{E_{nn}}\right)^{-1}
,\label{tau}\\  
{\cal G}_1(q,k;S_{2n})&\equiv&\int_{-1}^1dy \frac{2A}{2A S_{2n}+{(A+1)}(q^2+k^2)+2{kqy}}
,\nonumber\\
{\cal G}_2(q,k;S_{2n})&\equiv&\int_{-1}^1dy \frac{2A}{2A S_{2n}+{2A}q^2+{(A+1)}k^2+2A{kqy}}
.\end{eqnarray}
The above set of coupled equations can also be derived from  a renormalized Hamiltonian as 
shown in \cite{FrePPNP12}, where the associated renormalization group properties are also discussed.  
The minus ($-$) sign refers to a bound state subsystem and the plus sign ($+$) to a virtual state 
subsystem. 
Therefore, within the perspective of a more general $\alpha-\alpha-\beta$ system, 
the following cases can be described by the above coupled integral equations:
{\it all-bound} configuration, when there is no unbound subsystems;
{\it Borromean} configuration, when all the subsystems are unbound; 
{\it tango} configuration~\cite{RobPRA1999,JenRMP2004,ZinJPG13}, when we 
have two unbound and one bound subsystems; and {\it samba} configuration~\cite{YamNPA04}, 
when just one of the two-body subsystems is unbound. In the present case, as we are concerned
with $n-n-c$ halo nuclei system, only {\it samba} and {\it Borromean} configurations are 
possible, once we take that $n-n$ is unbound with a virtual-state energy of about $143$ keV.
This implies that only the sign $+$ is to be considered for $\tau_{nn}$ in Eq.~(\ref{tau}).

One can further simplify Eq.~(\ref{spec}), for numerical purpose, by having an uncoupled 
integral equation for $\chi_{nc}$:
{\small 
\begin{eqnarray}
\label{specnc}
 \chi_{nc}({q})&=&\tau_{nc}(q;S_{2n})\int_0^\infty k^2dk \;{\cal K}_\mu(q,k;S_{2n})\chi_{nc}(k),
 \nonumber\\
{\cal K}_\mu(q,k;S_{2n})&\equiv&
 \left[ {\cal G}_1(q,k;S_{2n}) - {\cal G}_1(q,k;\mu^2) \right] + 2\int_0^\infty p^2dp\;\tau_{nn}(p;S_{2n})\times 
\\ 
&\times&
\left[ {\cal G}_2(q,p;S_{2n}) - {\cal G}_2(q,p;\mu^2) \right]\left[ {\cal G}_2(k,p;S_{2n}) - {\cal G}_2(k,p;\mu^2) \right]
.\nonumber \end{eqnarray} 
}
The corresponding $s-$wave three-body wave-function can be written in terms of the spectator
functions  $\chi_{nn}(q)$ and $\chi_{nc}(q)$ as:
\begin{eqnarray}
\langle\vec{q}_c\vec{p}_c|\Psi\rangle&=&
\frac{\chi_{nn}(q_c)+\chi_{nc}(|\vec{p}_c-\frac{\vec{q}_c}{2}|)
+\chi_{nc}(|\vec{p}_c+\frac{\vec{q}_c}{2}|)}{S_{2n}+p_c^2+\frac{A+2}{4A} q_c^2},
\label{psiqc}
\end{eqnarray}
where $\{ |\vec{q}_c\vec{p}_c\rangle\}$ is the relative Jacobi momentum basis,  with $\vec{q}_c$ the relative 
momentum of the core to the center-of-mass of the $n-n$ system, and $\vec p_c$ the relative momentum
between the two neutrons. 
Note that, as we are going to present results corresponding to the limit cycle, namely, when all 
involved energies tends to zero with respect to the subtraction or regularization scale, we have dropped 
the regularization term in the denominator of the wave-function, which was introduced in Ref.~\cite{YamNPA04}. 
The configuration space halo wave-function, which is given by the Fourier Transform of the momentum 
wave-function, is an eigenstate of the free Hamiltonian, except when two particles are at the same point, 
such that in our model the two halo-neutrons are always found in the classically forbidden region. 
This model can represent a real halo state as long as the neutrons have a large probability to be found outside 
of the potential range and of the core. 

From the wave-function, given in momentum space by Eq.~(\ref{psiqc}), we can define the 
core momentum distribution for the $n-n-c$ system as
\begin{equation}
\label{nqc}
n(q_c)=\int d^3p_c |\langle\vec{q}_c\vec{p}_c|\Psi\rangle|^2, 
\end{equation}
with normalization such that $\int d^3q_c\, n(q_c)=1$.
In the context of cold atoms the large momentum behaviour of the above momentum density has been studied  
in detail for three-bosons in ~\cite{castin2011} and for mass imbalanced systems in \cite{YamPRA13}. The 
log-periodic solution of the spectator equations (\ref{spec}) in the ultraviolet limit, when $\mu\to \infty$, is 
the key property to derive asymptotic formulas for the one-body momentum densities. Furthermore,
it was verified in \cite{YamPRA13} how the solutions of (\ref{spec}) approaches the log-periodic form for the 
higher Efimov excited states. In addition, it was shown that the density properties at low momentum behaviour 
are universal, namely, approach the limit-cycle already for the ground state with finite $\mu$, and 
depend on the three-body binding energy and scattering lengths. 

\section{Results and discussion}

\begin{figure}[tbh!]
\begin{center}
\includegraphics[scale=0.6,clip]{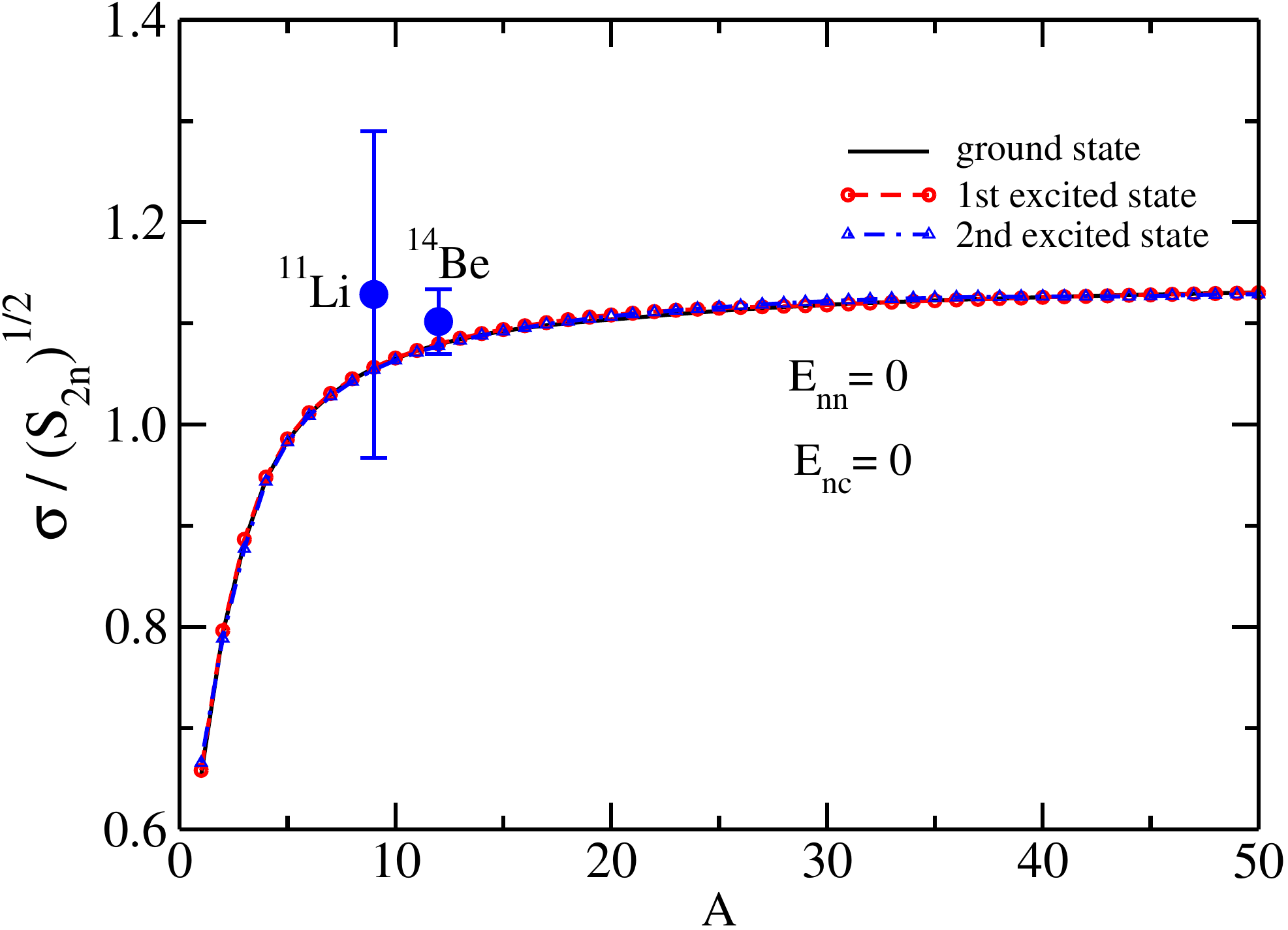}
\end{center}
\caption{Scaling plot for the core recoil momentum distribution $\sigma$ in the Efimov limit as a function of 
the core mass number $A$.  Experimental widths are from Refs.~\cite{TanJPG96} and \cite{ZahPRC93}, 
for $^{11}$Li and $^{14}$Be, respectively.} 
\label{scalsigma}
\end{figure}

The solution for the set of integral equations (\ref{spec}) provides the spectator 
functions and ultimately the momentum probability density (\ref{nqc}). We start by showing results for
$E_{nn}=E_{nc}=0$, in order to study the limit-cycle for the core momentum distribution in the
context of the two-neutron halo nuclei. To illustrate this limit we show in Fig.~\ref{scalsigma}
the corresponding scaling function (\ref{sigmac}) for $\sigma$, in terms of the dimensionless 
ratio $\sigma/\sqrt{S_{2n}}$ as a function of the core mass number. 
Results for the ground and two excited states in Fig. \ref{scalsigma} show that the limit-cycle is universal and in practice found for the ground state.
We compare with the experimental values of 
$\sigma/\sqrt{S_{2n}}$ obtained for  $^{11}$Li, $\sigma=21(3)$MeV/c, coming 
from the halo breakup reaction $^{11}$Li +C$\,\to\,^9$Li + X at 800 MeV/A~\cite{TanJPG96}, 
and for $^{14}$Be, which has a FWHM$=\, 92.7\,\pm \, 2.7$MeV/c  
for the core recoil momentum distribution~\cite{ZahPRC93} and $S_{2n}=$1.337 MeV~\cite{AudNPA03}.
The flattening of the scaling function for large $A$ reaching an asymptotic value 
can  be understood by inspecting the set of coupled equations (\ref{spec}) and the wave-function
(\ref{psiqc}) by noticing that the limit $A\to\infty$ can be performed, where all dependences on
$A$ are cancelled out. 
One has to remind that even for $A\to\infty$ the dependence of the core momentum distribution on
 the relative momentum  $q_c$ just reflects the momentum distribution of the center of mass 
 the two halo-neutrons in the nucleus. On the other hand, for $A\to 0$, the momentum distribution 
 tends be concentrated at small momentum as one can easily check that the relevant contribution to the
 integral equation for the spectator function comes from small momentum and $\sigma\to 0$. Naively, the light particle explores large distances, as the characteristic momentum is of the order of 
 $\sqrt{S_{2n}/A}$, and therefore $\sigma \sim \sqrt{A}$ for $A\to 0$. 

\begin{figure}[tbh!]
\begin{center}
\includegraphics[scale=0.32,clip]{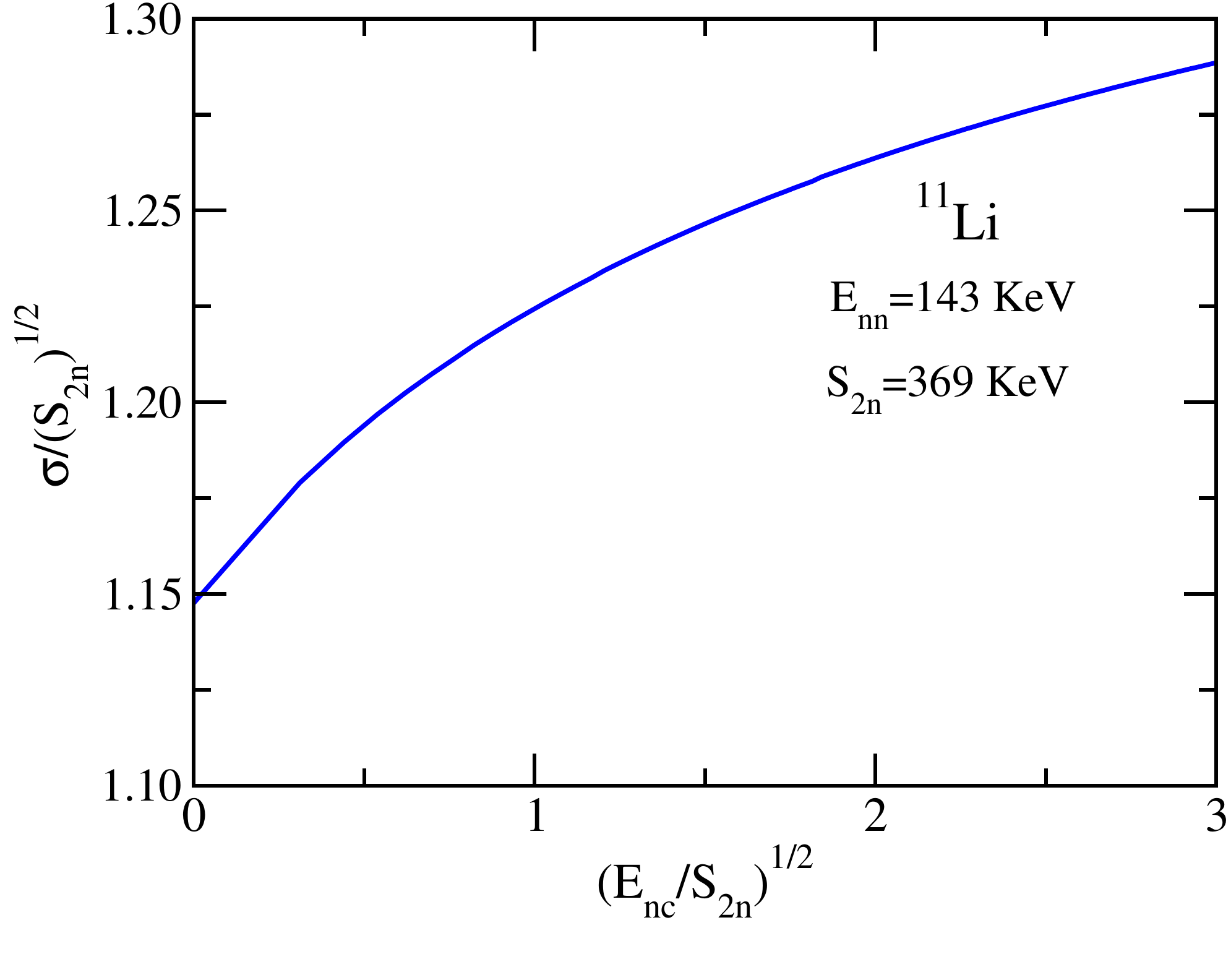}
\hspace{0.15cm}
\includegraphics[scale=0.32,clip]{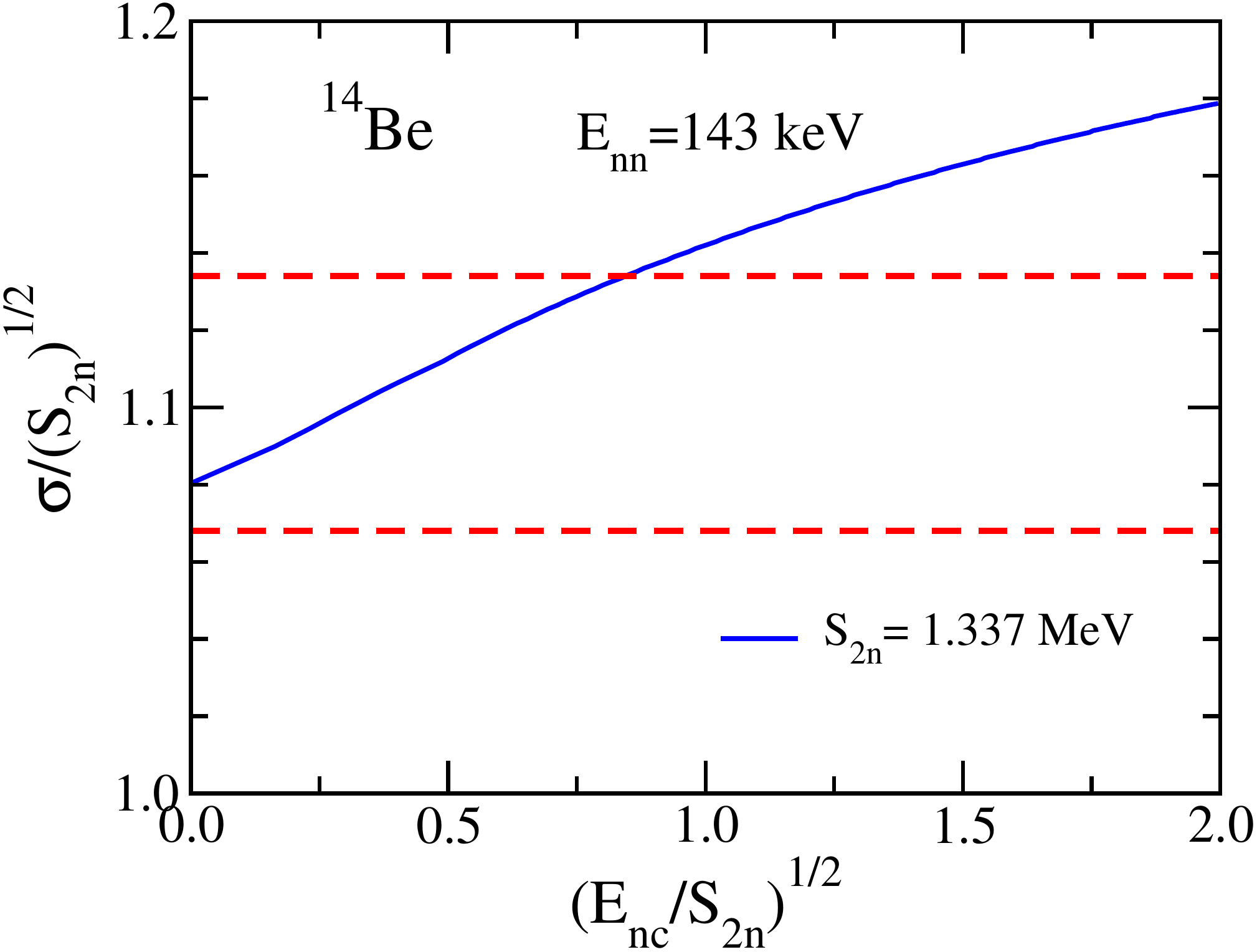}
\end{center}
\caption{Scaling plots for the core momentum distribution $\sigma$ in $^{11}$Li {\it (left-frame)} 
and $^{14}$Be {\it (right-frame)} for the fixed $E_{nn}=$ 143 keV (virtual energy). 
The experimental $S_{2n}=$ 369 keV \cite{SmiPRL08}  and 1.337 MeV~\cite{AudNPA03}, for
$^{11}$Li and $^{14}$Be, respectively. 
The dashed lines for $^{14}$Be  represent the region delimited by the experimental value 
FWHM$=\, 92.7\,\pm \, 2.7$MeV/c~\cite{ZahPRC93}.} \label{sigli11be14}\end{figure}

\begin{figure}[tbh!]
\includegraphics[scale=0.33,clip]{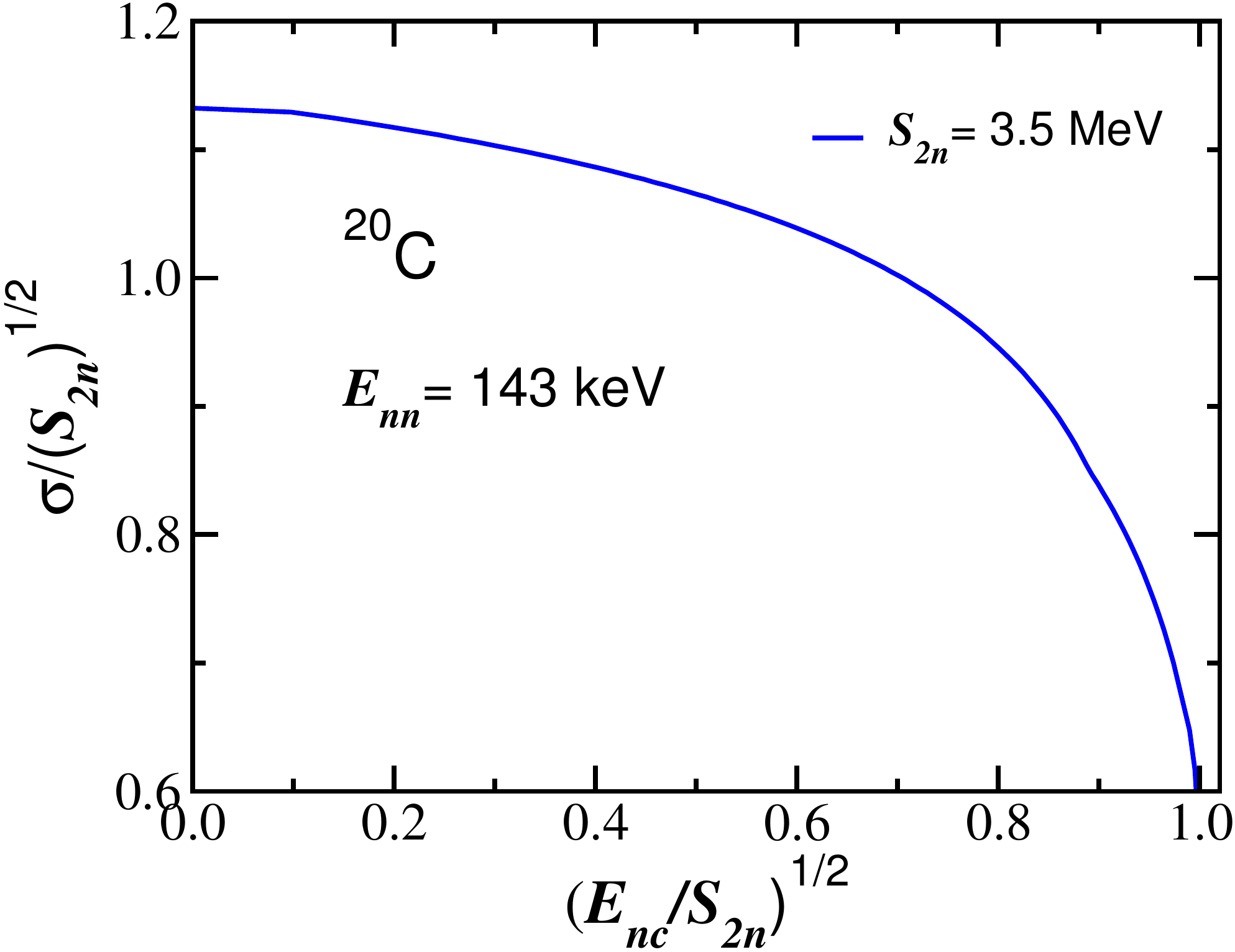}\hspace{0.1cm}
\includegraphics[scale=0.33,clip]{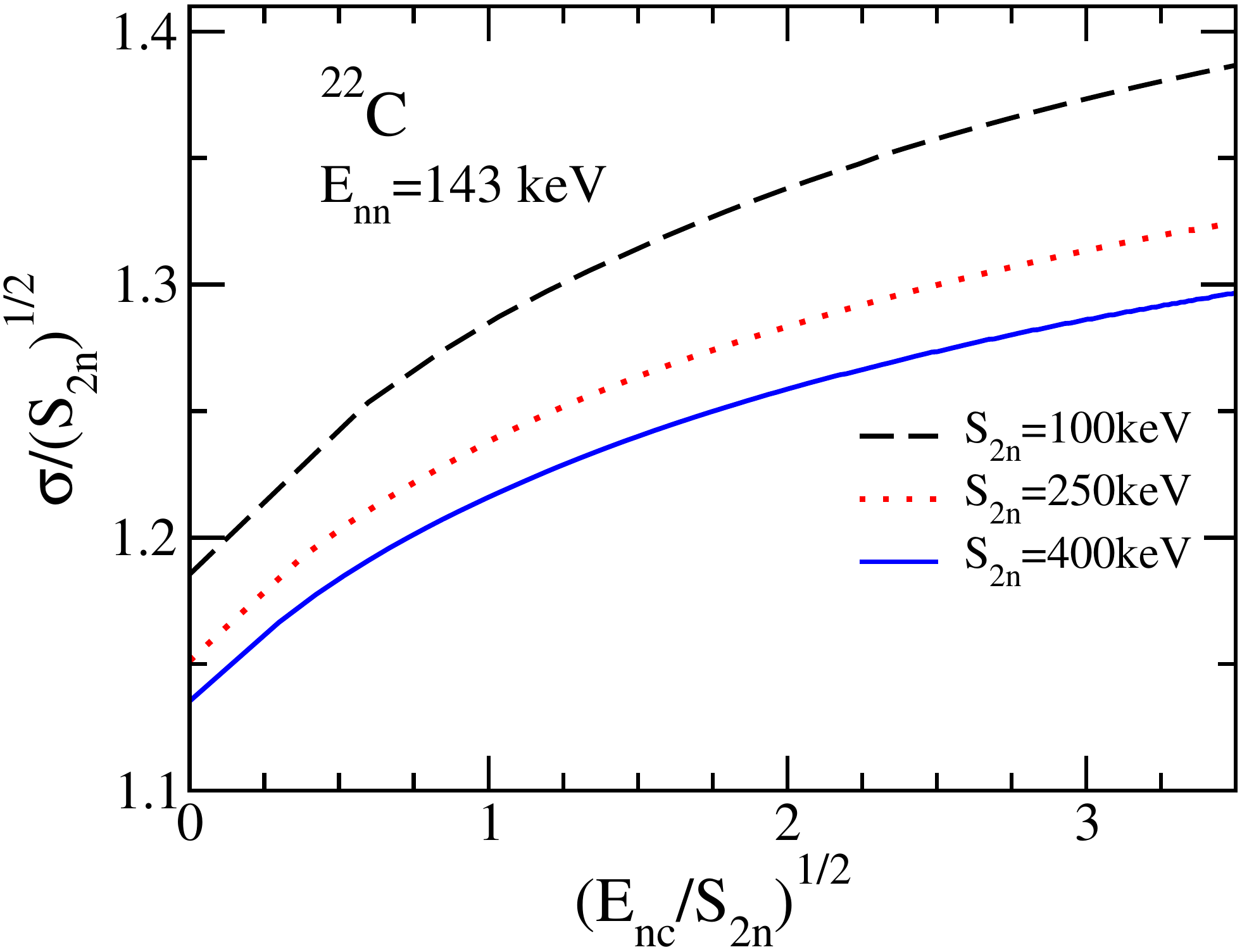}
\caption{Scaling plots for the core momentum distribution $\sigma$ for $^{20}$C {\it (left-frame)} 
and $^{22}$C {\it (right-frame)}, for a fixed $E_{nn}=143$ keV (virtual-state energy).
In the left frame, for $^{20}$C, we use $S_{2n}=$ 3.5 MeV~\cite{AudNPA03}. 
In the right frame, for $^{22}$C, we use three values for $S_{2n}$: 
100 keV (dashed line), 250 keV (dotted line) and 400 keV (solid line).} 
\label{sigc20c22}
\end{figure}

The dependence of $\sigma$ on the subsystems energies $E_{nn}$ and $E_{nc}$ is investigated
by considering the results presented in Figs. \ref{sigli11be14} and \ref{sigc20c22}, where we 
consider the Borromean configurations, in the cases of $^{11}$Li, $^{14}$Be and $^{22}$C, as 
well as the {\it samba}  type configuration (bound $nc$ subsystem), which is exemplified by the 
case of  $^{20}$C. For $E_{nc}=0$ and by changing $E_{nn}$ from 0 to 143 keV, $\sigma$ 
increases with respect to $\sqrt{S_{2n}}$, as is seen when the values at the origin of these 
figures are compared to Fig.~\ref{scalsigma}. 
It means that the halo shrinks as the virtual state energy increases in absolute value. This effect
was found in \cite{YamNPA04}, namely for a given $S_{2n}$ the size of the halo
shrinks when going from {\it all-bound} configuration to the {\it Borromean} one. This behaviour happens
because the interaction becomes less attractive, such that to keep the three-body binding energy 
the state has to become smaller. This effect is also observed as the value of $E_{nc}$ increases
in for  $^{11}$Li {(\it left-frame)} and $^{14}$Be {(\it right-frame)}, 
as shown in Fig.~\ref{sigli11be14}. For the $s-$wave virtual state energy
of $^{10}$Li of 50 keV, one has $\sigma/\sqrt{S_{2n}}=1.18$ and $\sigma=22$ MeV/c compared to 
the experimental value of $\sigma=21(3)$MeV/c~\cite{TanJPG96}. The experimental value of 
$\sigma=\,39.4\,\pm\, 1.1$ MeV/c from the FWHM of the momentum distribution of $^{14}$Be 
\cite{ZahPRC93}
is represented by the region delimited with the dashed lines in the right-frame 
of Fig.~\ref{sigli11be14}, and from that we could say roughly that $s-$wave virtual state of
$^{13}$Be $E_{nc}$ is less than 1 MeV, which is consistent with 0.2 MeV that is the known 
value (see e.g. \cite{YamNPA04}). 
We note that the dependence on $E_{nc}$ is very mild and by changing it
from 0 to $S_{2n}$ a variation of $\sigma/\sqrt{S_{2n}}$ of only 10\% is 
found in our model, which puts a constraint in the error in the experimental 
ratio $\sigma/\sqrt{S_{2n}}$ in order to be useful to extract
information on the neutron-core virtual state energy. 

In Fig.~\ref{sigc20c22}, we show results for the scaling plots for the core momentum distribution $\sigma$
in $^{20}$C  (left-frame) and $^{22}$C (right-frame). In the left frame, the subsystem $n-^{18}$C forming 
the $s-$wave one neutron halo $^{19}$C is bound with
{energy $E_{nc}\equiv S_{1n}=$ 580 keV~\cite{AudNPA03}, where $S_{1n}$ is the}  
one neutron separation energy.
 Although, all halo low-energy scales are known for 
$^{20}$C, we allow variation of the ratio $E_{nc}/{S_{2n}}$ to illustrate how the 
width of the momentum 
distribution varies in the case of halo nuclei with bound $n-c$ subsystem.
The width decreases as $E_{nc}/{S_{2n}}$ increases as the bound state energy becomes
closer to the lowest scattering threshold, and consequently the neutron distance
to core increases leading to the sudden drop of $\sigma$ to zero, when $E_{nc}/S_{2n}$ goes to
unity. In the right-frame of the figure, we present results for $\sigma/\sqrt{S_{2n}}$ 
as a function of $(E_{nc}/{S_{2n}})^{1/2}$ for
 $^{22}$C computed with different values of $S_{2n}$, 100 keV, 250 keV  and 400 keV. 
 {We observe in the figure that while $\sigma$ } exhibits a strong dependence 
on $S_{2n}$ with $E_{nn}$ and $E_{nc}$ 
kept constant, the variation of $\sigma$ with the ratio 
$E_{nc}/{S_{2n}}$ for $S_{2n}$ constant shows a quite weak sensitivity, 
as one could expect for the Borromean case. In that sense, as already recognized, the value of 
$\sigma$ gives a good constraint for $S_{2n}$ in this case.

\begin{figure}[tbh!]
\includegraphics[scale=0.33,clip]{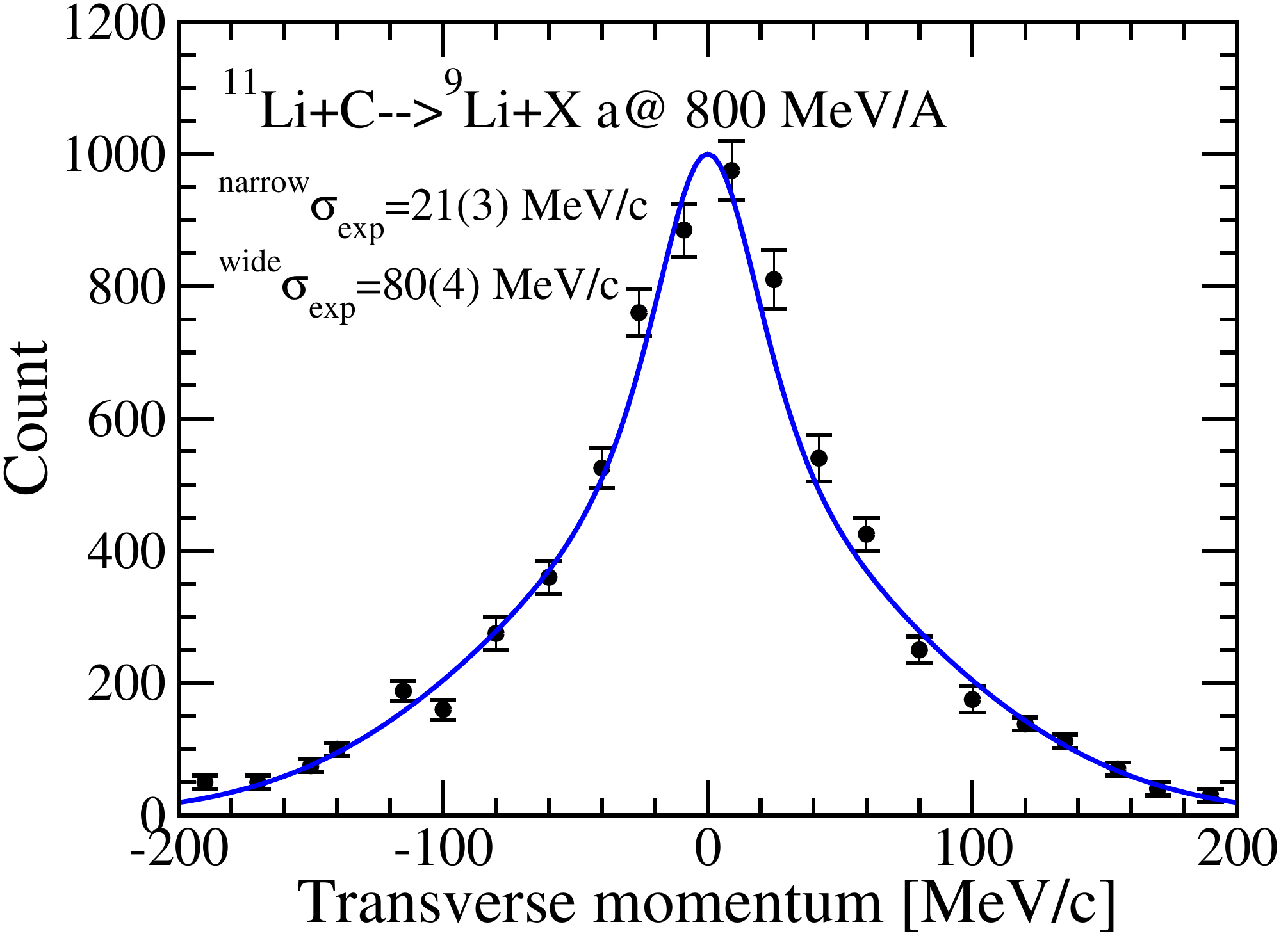}\hspace{0.2cm}
\includegraphics[scale=0.33,clip]{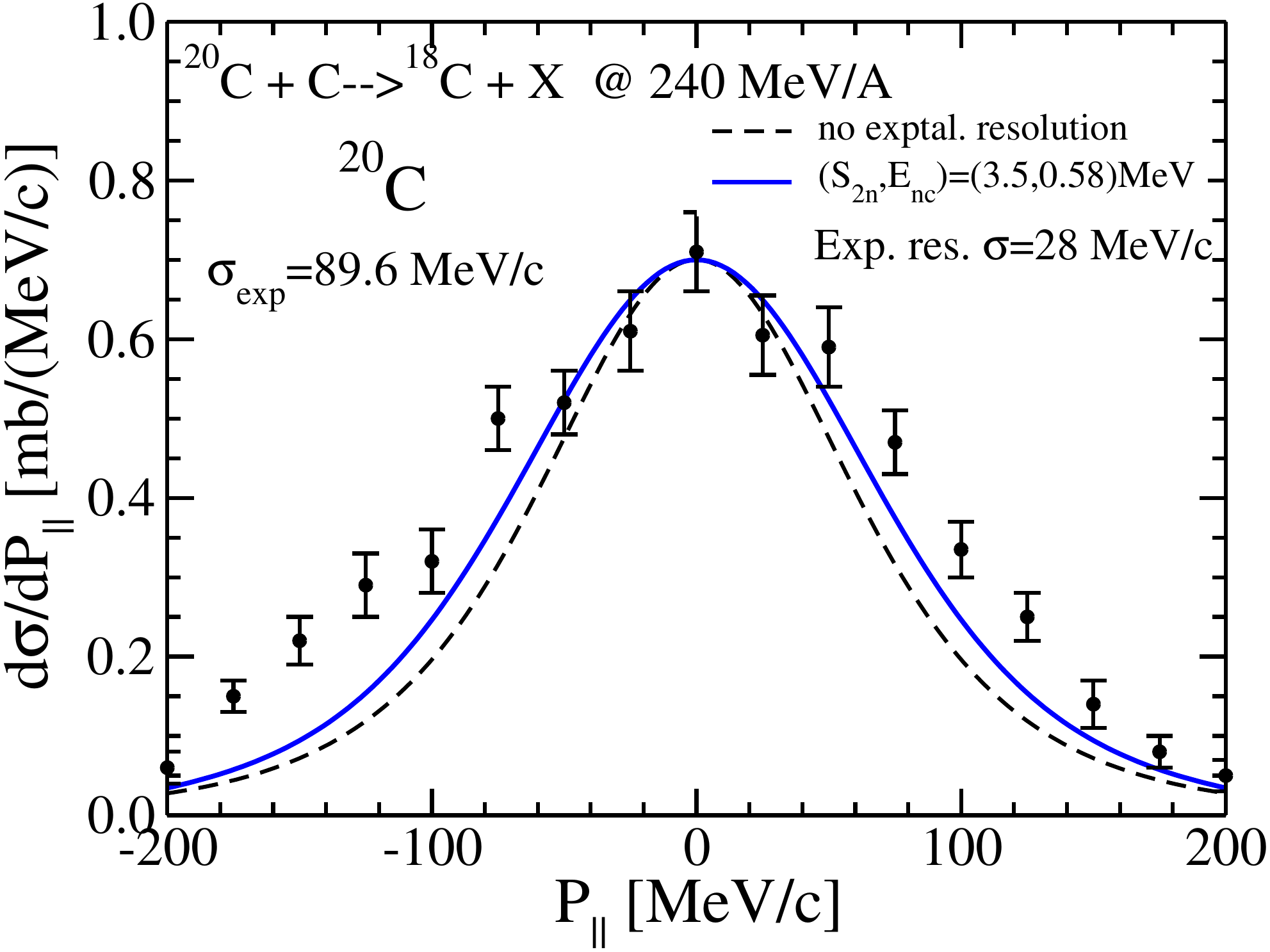}
\caption{Distributions of the recoil core in $^{11}$Li {\it (left-frame)} and 
$^{20}$C {\it (right-frame)} observed in the halo breakup reaction
 in a target compared to our
calculations of the momentum distribution normalized to the data.   
The narrow distribution 
for $^{11}$Li is computed with  $S_{2n}=\,$369 keV \cite{SmiPRL08}, $s-$wave virtual state 
energy of $^{10} $Li, $E_{nc}$=50 keV, and singlet $n-n$ virtual state, $E_{nn}$=143 keV. 
The results for the  distribution with
the computed narrow $\sigma=$ 22 MeV/c ($\sigma_{exp}=$ 21(3) MeV/c) are added to a wide one with 
$\sigma_{exp}=$ 80 MeV/c. 
The calculations were performed for the experimental values from ~\cite{AudNPA03} 
of $S_{2n}=$ 3.5 MeV for $^{20}$C, $S_{1n}=$ 580 keV for $^{19}$C. 
The experimental results for $^{11}$Li are extracted from \cite{TanJPG96} and for $^{20}$C
from \cite{KobPRC12}. For $^{11}$Li the experiment detected the $^{9}$Li  
transverse momentum to the beam and for
$^{20}$C the inclusive parallel momentum of $^{18}$C.
The distribution for $^{20}$C was 
folded with the experimental resolution of $\sigma=28$ MeV/c.} \label{distli11c20}
\end{figure}


\begin{figure}[tbh!]
\includegraphics[scale=0.33,clip]{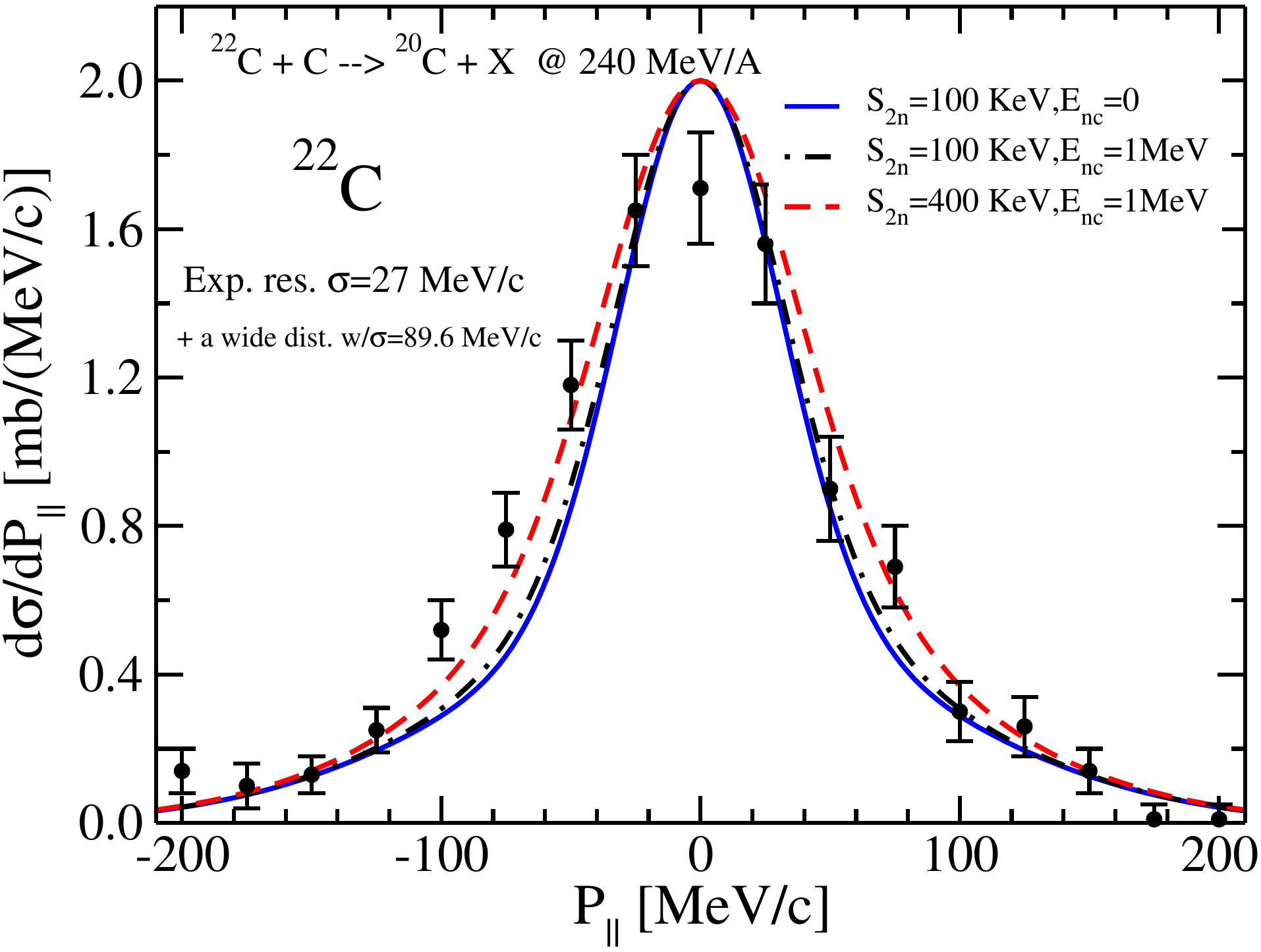}\hspace{0.2cm}
\includegraphics[scale=0.33,clip]{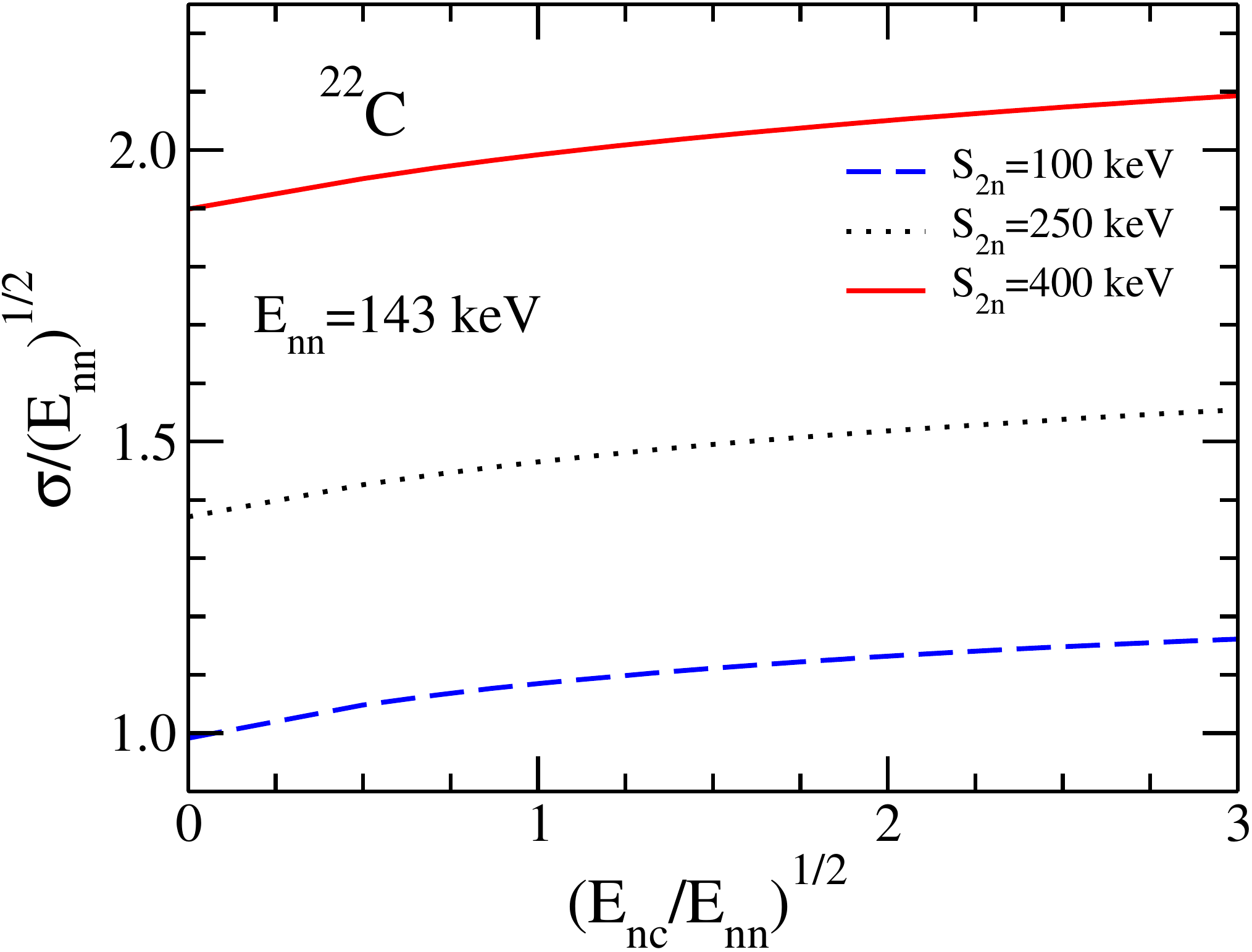}
\caption{{\it Left-frame:} Distribution of the recoil core $^{22}$C obtained with the zero-range model 
compared to the experimental data from \cite{KobPRC12} for different inputs. 
For $^{22}$C the experiment detected the inclusive parallel momentum of $^{20}$C.
Results for
$\left(S_{2n}[keV],E_{nc}[MeV]\right)$ folded with the experimental resolution of $\sigma=27$ MeV/c and added 
to a wide distribution with $\sigma=89.6$~MeV/c: 
solid-line (100,0), dashed-line (400,1) and dotted-line (100,1).  
{\it Right-frame:} Scaling plot for the core momentum width ($\sigma$) in $^{22}$C 
(right frame) for given singlet $n-n$ virtual state energy (143 keV) and $S_{2n}$ of 
100 keV (dashed line), 250 keV (dotted line) and 400 keV (solid line).
} \label{distc22}
\end{figure}

After our discussion of the general scaling properties of the with of the  
momentum distribution, we show in  Fig. \ref{distli11c20} 
our calculations of the core recoil momentum distribution for 
$^{11}$Li (left-frame) and $^{20}$C (right-frame) compared to actual 
results from halo breakup experiments obtained reactions with carbon target at 
800 MeV/A \cite{TanJPG96} and at 240 MeV/A \cite{KobPRC12}, respectively. 
For $^{11}$Li, a wide distribution with $\sigma=$80 MeV/c is added to the computed 
narrow one, which has $\sigma=$ 22 MeV/c. We remark that all three inputs to compute 
the narrow distribution are fixed to known values of $S_{2n}=\,$369 keV~\cite{SmiPRL08}, 
the $s-$wave virtual state energy of $^{10}$Li, $E_{nc}$=50 keV, and the  
singlet $n-n$ virtual state, $E_{nn}$=143 keV. 
The wide momentum distribution is beyond our model, which is more concerned on the halo 
neutrons. That contribution should be associated with inner part of the halo neutron orbits, 
close to the core region. In the comparison with the experimental data, the normalisations of 
the wide and narrow distributions are fitted to the data.  After that, we find a fair reproduction 
of the experimental data as shown in the figure. This procedure confirms that our approach is 
a viable tool to extract information on the large two-neutron halo properties from the core 
momentum distribution.

The right-frame of Fig.~\ref{distli11c20} presents the core momentum distribution for $^{20}$C. 
The calculations were performed with $S_{2n}=$ 3.5 MeV and with $^{19}$C one-neutron 
separation energy equal to 580 keV~\cite{AudNPA03}.
The model is compared to data obtained from \cite{KobPRC12}, after folding
with the experimental resolution of $\sigma=28$ MeV/c. We observe that a wide distribution is 
somewhat missing to fit the experimental results in this case.

The model results for the core recoil momentum distribution in $^{22}$C, with two-neutron 
separation energies of {100 and 400 KeV,} is presented in the left-frame of Fig.~\ref{distc22}, 
and compared to data obtained from \cite{KobPRC12}.  The singlet virtual $n-n$ energy is fixed 
to 143 keV, with the virtual-state energy of $n-^{20}$C chosen as 0 and 1 MeV~\cite{MosNPA13}. 
The narrow theoretical distribution is folded to the experimental resolution of $\sigma=27$ MeV/c 
and added to a wide one with $\sigma=89.6$~MeV/c.
The results presented in this figure illustrate the weak sensitivity of the core recoil momentum 
distribution to the variation of the virtual-state energy of $^{21}$C, which is taken between 0 and 
1 MeV, as it was shown by the results with $S_{2n}=$ 100 keV. The difference between 
the distributions obtained with $S_{2n}=$ 100  and 400 keV, computed with $E_{nc}=1$ MeV
is not enough to discriminate $S_{2n}$ in view of the experimental data error. The model
sensitivity to the physical inputs, in the interesting case of $^{22}$C is further explored,
in the right-frame of the figure,  where the scaling plot 
{for $\sigma/\sqrt{E_{nn}}$ as a function of $\sqrt{E_{nc}/E_{nn}}$} 
is shown for three values of $S_{2n}$. The weak 
sensitivity to the $s-$wave virtual state energy of $^{21}$C is seen and one could consider 
to obtain an upper limit to the experimental relative error in order to extract information 
on the $n-^{20}$C scattering length. However, a variation of the ratio $E_{nc}/E_{nn}$ 
between 0 and 9 gives 10\% variation of $\sigma$ ($E_{nn}$ is fixed), which is surmounted by  
a variation of about 50\% in $S_{2n}$. Therefore, it is required an independent source of information 
to constrain the $n-^{20}$C scattering length, which we can find from the matter radius of $^{22}$C \cite{TanPRL10}.

\begin{figure}[tbh!]
\vspace{-0.cm}
\centerline{\includegraphics[width=14cm,clip]{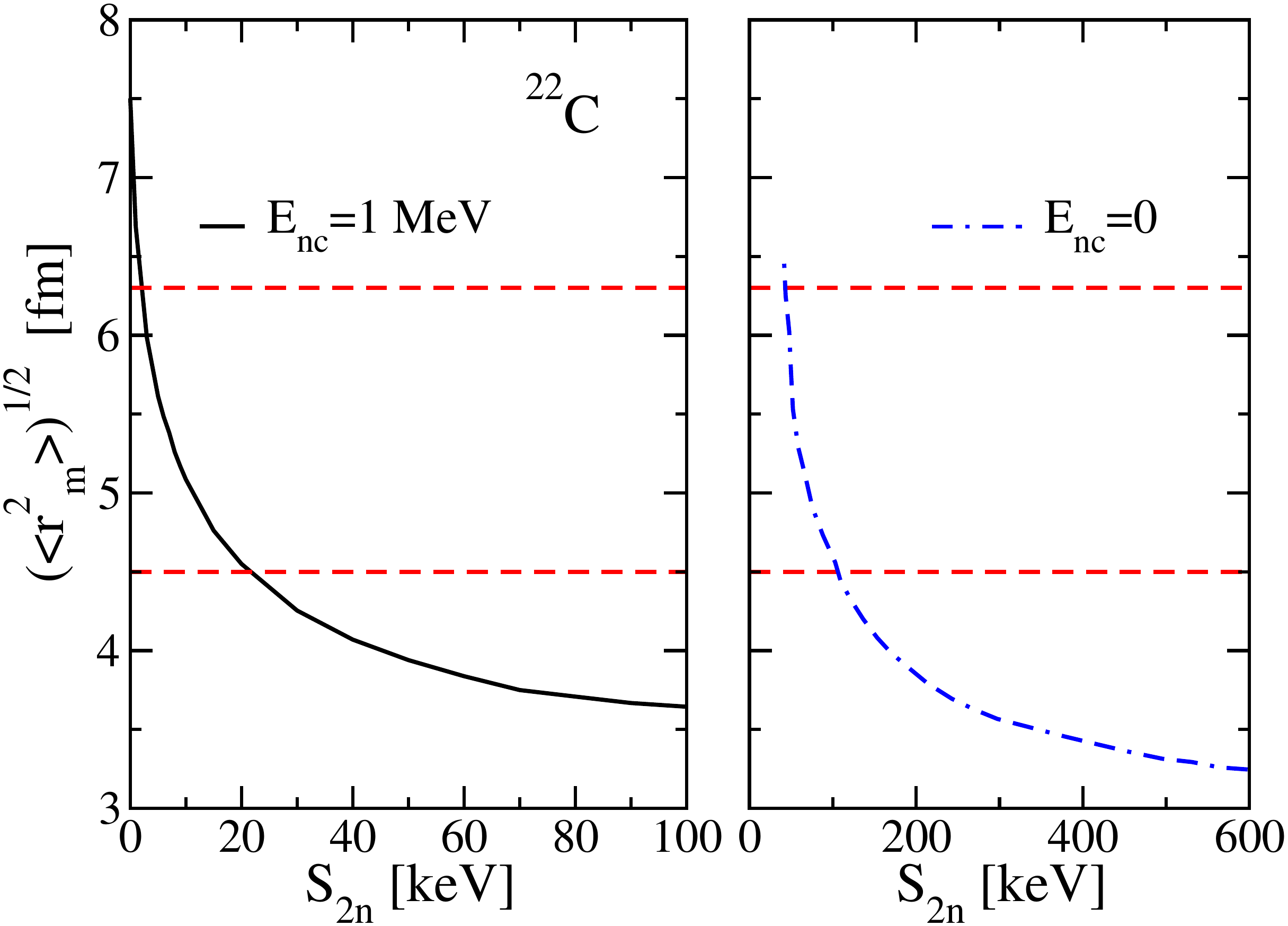}}
\caption{Root-mean-square (rms) matter radius of $^{22}$C given as a function of the two-neutron 
separation energy, computed with the singlet $n-n$ virtual-state energy fixed ($E_{nn}=$143 keV), 
considering the virtual-state energy of $n-^{20}$C given by $E_{nc}=$
 1 MeV (left frame) and 0 MeV (right frame). 
The dashed lines represent the upper and lower limits for the experimental value 
$5.4\pm0.9$~fm reported in \cite{TanPRL10}.
} \label{rn}
\end{figure}

To close our discussion of $^{22}$C, we computed the matter radius starting with the rms  
radius of the halo neutrons $(r_n)$ with respect to the center-of-mass, which is obtained
from the configuration space $n-n-c$ wave-function, which is obtained by considering the 
Fourier transform of the corresponding momentum wave-function (\ref{psiqc}). For details
on this procedure, see \cite{YamNPA04,YamPLB11}.  
The corresponding formula of the matter radius is given by  
$\sqrt{\langle r^2_m[^{22}C]\rangle}=\sqrt{(2/22)\,\langle r^2_n\rangle\,+\,(20/22)\,\langle r^2_m[^{20}C]}\rangle$. 
The $^{20}$C matter radius is $r_m[^{20}C]=$ 2.98(5) fm\cite{OzaNPA01}. 
The plot of Fig.~\ref{rn} shows the theoretical values of the rms matter
radius of $^{22}$C as a function of $S_{2n}$ for a fixed $s-$wave virtual 
state energies of the singlet $n-n$ pair (143 keV) and $^{21}$C (1 MeV 
\cite{MosNPA13} and 0) compared to data from \cite{TanPRL10}. 
In the figure, the  limits for the extracted matter radius~\cite{TanPRL10} are shown, 
and we can make some remarks analysing the consistence between the different 
available data and our model, considering that 100 keV $\lesssim S_{2n} \lesssim 400$ keV: 
(i)  for $E_{nc}=0$, one finds that 3.5 fm $\lesssim \langle r^2_m[^{22}C]\rangle^{1/2} \lesssim $ 4.5 fm;
and, (ii) for $E_{nc}= 1$ MeV, we have $\langle r^2_m[^{22}C]\rangle^{1/2} \lesssim $ 3.5 fm. 
Only if $E_{nc}\sim 0$ we obtain a region for $ S_{2n}$ close to 100 keV, consistent with rms 
matter radius of $^{22}$C within one standard deviation  and in the lower bound of the radius, namely $\sim 4.5$ fm.
The value of $E_{nc}= 1$ MeV for the virtual state of $^{21}$C and $S_{2n}\sim$ 100 keV
is compatible with two standard deviation; from that, $ \langle r^2_m[^{22}C]\rangle^{1/2} \lesssim $ 3.5 fm. 
This combined analysis for $^{22}$C of the core recoil momentum distribution, with rms matter radius
and virtual state energy of $n-^{20}$C, suggests that such virtual-state energy and matter radius are 
overestimated. Independent new data on the $S_{2n}$ for $^{22}$C could help in clarifying the 
tension between data analysis with the present universal model.  

\section{Conclusions}
In summary, by considering the renormalized zero-range model applied to the case of  core recoil 
momentum distributions of  $^{11}$Li and $^{14}$Be, we found a fair consistency with experimental 
data, just by using the known low-energy parameters. Relying on the fact that such simplified model 
gives already a valid description of the two-neutron $s$-wave halo, we proceed with a combined 
analysis of recent experimental data on the core momentum distribution in $^{22}$C, which is given 
by Kobaiashi et al.~\cite{KobPRC12}, the corresponding rms matter radius and the  
$^{21}$C virtual state energy. Our conclusion is that, with the value of the two-neutron separation energy 
of $^{22}$C given in the interval from 100 to 400 keV, the rms matter radius of $^{22}$C will be
within two standard deviations if the virtual state energy of $^{21}$C is close to 0. By considering the 
$^{21}$C with a virtual-state energy between 0 and 1 MeV, the matter rms radius should be 
between 3.5 and 4.5 fm. To reconcile a virtual-state energy with $E_{nc}\sim$1 MeV, a matter radius 
of 5.4$\pm$0.9 fm and 100 keV $\lesssim S_{2n} \lesssim 400$ keV, the possibility is 
$S_{2n}\sim$ 100 keV and $E_{nc}<$ 1 MeV, implying that 
$ \langle r^2_m[^{22}C]\rangle^{1/2} \sim $ 4.5 fm. 
 A refined analysis of the core momentum distribution, beyond  the Serber model~\cite{Serber}, is 
 desirable, of course. However, the comparison of results obtained by the present model for
 $^{11}$Li and $^{14}$Be with corresponding data suggests small corrections to the distribution
 verified for $^{22}$C. 

We thank partial support from the Brazilian agencies FAPESP, CNPq and CAPES.


\vspace{-.5cm}
\begin{thebibliography}{18}
\vspace{-0.5cm}
\bibitem{TanJPG96} I.~Tanihata, J. Phys. G {\bf 22} (1996) 157.
\bibitem{RiiPST13} K. Riisager, Phys. Scr. T\textbf{152} (2013) 014001. 
\bibitem{Efimov70}
V. Efimov, Phys. Lett. B {\bf  33} (1970) 563.

\bibitem{Efimov11}
V. Efimov, Few-Body Syst. {\bf 51} (2011) 79.

\bibitem{Ferlaino11}
F. Ferlaino, A. Zenesini, M. Berninger, B. Huang, H.-C. N\"agerl, and R. Grimm,
Few-Body Syst. {\bf 51} (2011) 113.

\bibitem{FreFBS14} T. Frederico, Few-Body Syst. {\bf 55} (2014) 651.

\bibitem{FrePPNP12} T. Frederico, M. T.  Yamashita,  A. Delfino, and L. Tomio, 
Prog. Part. Nucl. Phys. {\bf  67} (2012) 939.

\bibitem{ZahPRC93} M. Zahar, et al., Phys. Rev. C {\bf 48} (1993) R1484.

\bibitem{KobPRC12} N.~Kobayashi et al., Phys. Rev. C {\bf 86} (2012) 054604.

\bibitem{TanPPNP12} I. Tanihata, H.  Savajols, and R. Kanungo, 
 Prog. Part. Nucl. Phys. {\bf  68} (2013) 215.
 
\bibitem{FreFBS11} T. Frederico, L. Tomio, A. Delfino, M. R. Hadizadeh, M. T.
Yamashita, Few-Body Syst. {\bf  51} (2011) 87.

\bibitem{YamPRA13} M. T. Yamashita, F. F. Bellotti, T. Frederico, D. V. Fedorov, A. S. Jensen, N. T. Zinner, 
Phys. Rev. A \textbf{87} (2013) 062702.

\bibitem{YamPRA02} M. T. Yamashita, T. Frederico, A.  Delfino, L.  Tomio, 
Phys. Rev. A {\bf 66} (2002) 052702.

\bibitem{SKT} G. A. Skorniakov, K. A. Ter-Martirosian.
Sov. Phys. JETP {\bf 4} (1957) 648.

\bibitem{AudNPA03} G. Audi, A. H. Wapstra, and C. Thibault, Nucl. Phys. A 729, 337 (2003).

\bibitem{GauPRL12} L. Gaudefroy et al., Phys. Rev. Lett. {\bf 109} (2012) 202503.

\bibitem{MosNPA13} S. Mosby  et al.,  Nucl. Phys. A {  909}, 69-78 (2013)

\bibitem{TanPRL10} K. Tanaka {\it et al.}, 
{Phys. Rev. Lett.} {   104} (2010) 062701. 

\bibitem{OzaNPA01} A. Ozawa et al., Nucl. Phys. A 691 (2001) 599.

\bibitem{YamPLB11} M. T. Yamashita, R. M. de Carvalho, T. Frederico, L. Tomio, 
Phys. Lett. B \textbf{697} (2011) 90; Phys. Lett. B \textbf{715} (2012) 282.

\bibitem{ForPRC12} H. T. Fortune, R. Sherr, Phys. Rev. C \textbf{85} (2012) 027303.

\bibitem{AchPLB13} B. Acharya, C. Ji, D. R. Phillips, Phys. Lett. B \textbf{723} (2013) 196. 

\bibitem{AchFB15} B. Acharya, D. R. Phillips, arXiv:1508.02697 [nucl-th]. Contribution to the
21st Conference in Few-Body Problems in Physics.

\bibitem{SmiPRL08} M. Smith et al., Phys. Rev. Lett. {\bf 101}   (2008) 202501.

\bibitem{MarPRC01} F. M. Marqu\'es,  et al.,
Phys. Rev. C {\bf 64} (2001) 061301.

\bibitem{PetNPA04} M. Petrascu, et al.,
Nucl. Phys. A {\bf  738} (2004) 503.

\bibitem{suzuki-2006} W. Horiuchi and Y. Suzuki, Phys. Rev. C {\bf 74} (2006) 034311.

\bibitem{MarPLB00} F. M. Marqu\'es,  et al.
Phys. Lett. B {\bf 476} (2000) 219.
 
 \bibitem{YamPRC05} M. T. Yamashita, T. Frederico, L. Tomio,
Phys. Rev. C {\bf 72} (2005) 011601(R).

\bibitem{AmoPRC97} A. E. A. Amorim, T. Frederico and L. Tomio,
Phys. Rev.  C {\bf 56} (1997) R2378.

\bibitem{YamNPA04} M. T. Yamashita,  L. Tomio, T.  Frederico, 
Nucl. Phys. A  {\bf 735} (2004) 40.

\bibitem{RobPRA1999} F. Robicheaux, Phys. Rev. A {\bf 60} (1999) 1706.

\bibitem{JenRMP2004} A. S. Jensen, K. Riisager, D.V. Fedorov, 
E. Garrido, { Rev. Mod. Phys.} {\bf   76} (2004) 215.

\bibitem{castin2011} Y. Castin and F. Werner, 
Phys. Rev. A {\bf 83} (2011) 063614.

\bibitem{ZinJPG13}  N. T. Zinner, A. S.  Jensen,   
J. Phys. G: Nucl. Part. Phys. {\bf  40} (2013) 053101.

\bibitem{Serber} R. Serber, Phys. Rev. {\bf 72} (1947)  1008.

\end{thebibliography}
\end{document}